\documentclass[aps,pra,twocolumn, notitlepage,10pts]{revtex4-1}
\usepackage{times,bbm,amsmath,amssymb}
\usepackage{hyperref}
\usepackage{color}
\usepackage{graphicx}
\usepackage{bm}
\usepackage{soul}
\usepackage{lipsum}
\usepackage{subfig}
\usepackage{orcidlink}

\begin{document}

\title{Parallel trusted node approach for satellite quantum key distribution}

\author{Gianluca De Santis\orcidlink{0009-0005-2393-6076}}
\author{Konstantin Kravtsov\orcidlink{0000-0003-4499-4089}}
\author{Sana Amairi-Pyka\orcidlink{0000-0001-9502-4176}}
\author{James A. Grieve\orcidlink{0000-0002-2800-8317}}

\affiliation{Quantum Research Centre, Technology Innovation Institute, PO Box 9639 Abu Dhabi, United Arab Emirates}

\begin{abstract}
Quantum key distribution (QKD) via satellite links is the only currently viable solution to create quantum-backed secure communication at a global scale. To achieve intercontinental coverage with available technology one must adopt a ``flying trusted node'' paradigm, in which users fully trust the satellite platform. Here, inspired by the concept of distributed secret sharing and the imminent projected launch of several QKD-equipped satellites, we propose a parallel trusted node approach, in which key distribution is mediated by several satellites in parallel. This has the effect of distributing the trust, removing single points of failure and reducing the necessary assumptions. In addition, we discuss the versatility that an optical ground station should provide to execute such a protocol and, in general, to be fully integrated into a multi-party global quantum network.

\end{abstract}

\maketitle

\section*{Introduction}
Quantum Key Distribution (QKD) is widely regarded as the most mature application of quantum communications, providing a pathway to inherently secure communications \cite{gisin2007quantum}. Since its invention in 1984, much research has been dedicated to both the advancement of theoretical foundations and the development of practical realizations of quantum secure communication \cite{pirandola2020advances}. Despite recent developments in the adoption and deployment of QKD networks \cite{stanley2022recent} global connectivity remains challenging due to the limitation imposed by distance. Signal attenuation in optical fibers and atmospheric turbulence in free-space channels are the primary factors constraining the transmission of quantum signals, which are fundamentally sensitive to losses \cite{pirandola2017fundamental}. The exponential attenuation of optical fibers results in an inter-node range limit of several hundred kilometers \cite{yin2016measurement,frohlich2017long,yuan201810}, while diffraction and turbulence further curtail the range for free-space links \cite{erven2008entangled,schmitt2007experimental}.

Several strategies have been proposed to address this challenge, including novel protocols in which key rate scales as the square root of the channel transmittance \cite{lucamarini2018overcoming}. However, even such efforts are insufficient to bridge global distances without the development of quantum repeaters and quantum memories. To date, the challenge of scaling QKD networks to longer distances has been tackled in all cases by adopting the so-called ``trusted node'' paradigm, in which many shorter point-to-point links are established between physically secured nodes. While classical key swapping mechanisms enable any two nodes on such a network to share a key pair \cite{mehic2020quantum}, the key material is encoded classically within each node, requiring users to trust the physical integrity of all facilities in the chain. Although this solution appears to be tolerated in many metropolitan or national-scale networks, it is more challenging in an international environment, where communications chains would span multiple jurisdictions.

In this context, satellite platforms emerge as a natural component in the envisaged quantum network architecture \cite{sidhu2021advances,bedington2017progress}. Since early proposals in 2008 \cite{armengol2008quantum}, several studies from academic groups throughout the world have been released \cite{bonato2009feasibility,khatri2021spooky,bourgoin2013comprehensive,kaushal2015free,bedington2016nanosatellite,sidhu2021advances}. The feasibility of establishing a quantum channel from low Earth orbit (LEO) \cite{toyoshima2009polarization}, mid Earth orbit (MEO) \cite{dequal2016experimental}, and geostationary orbit (GEO) \cite{gunthner2017quantum} satellites, has been proved experimentally. Sources of entangled photons have been demonstrated on orbit \cite{tang2016photon} and entanglement distribution has been achieved from LEO \cite{yin2017satellite}. Finally, QKD protocols have been implemented via satellites \cite{takenaka2017satellite,yin2017satelliteQKD,liao2017satellite,liao2017space,yin2020entanglement}, and an integrated space-to-ground quantum network has been accomplished with 150 users \cite{chen2021integrated}.

Among the array of potential configurations, a flying trusted node (TN) approach has enjoyed much focus in recent years since it does not rely on quantum memories \cite{wittig2017}, nor the simultaneous transmission of a pair of photons to separate receivers \cite{yin2017satellite}.
At the present stage of technology development, a variety of satellite QKD hardware and software solutions have been proposed. Despite an absence of consensus on implementation details, considerable effort is concentrated on demonstrating the feasibility and efficacy of satellite QKD. In this study, we propose a parallel trusted node approach such that the diversification of trust into different operators can give the users a practical advantage over possible eavesdropping or willful deception. This strategy places certain requirements on the ground stations to be used, and we discuss the main characteristics of versatility and interoperability of an optical ground station for quantum communication purposes. This proposal aims to further lower barriers to adoption of global satellite quantum communications with present technology.

\section{Satellite-based QKD}

Realization of a satellite QKD protocol offers various possible configurations. If the source resides on the satellite and the receiver is placed at the Optical Ground Station (OGS), the established link is termed a downlink; if the source is on the ground, the link is termed an uplink. The downlink configuration holds distinct advantages, primarily due to atmospheric turbulence affecting the latter segment of the link, resulting in reduced divergence compared to the uplink scenario where turbulence impacts radiation in the initial segment of the channel. Our proposal is insensitive to this distinction, and in fact can be adopted for an arbitrary combination of these scenarios.

\begin{figure*}
    \centering
    \subfloat[][\emph{}\label{sat_conf1}]
    {\includegraphics[height=.21\textwidth]{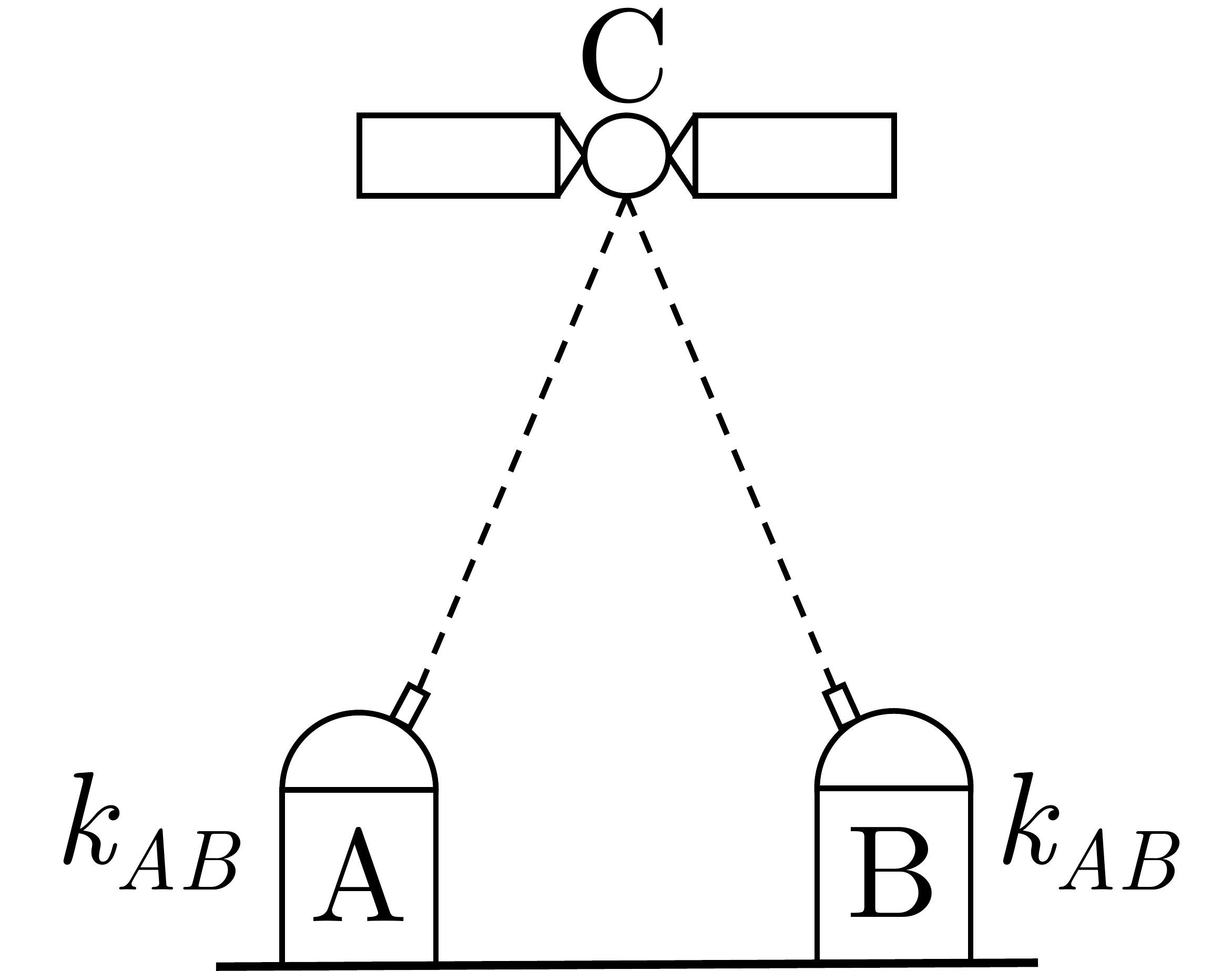}} \qquad \qquad \quad \quad
    \subfloat[][\emph{}\label{sat_conf2}]
    {\includegraphics[height=.25\textwidth]{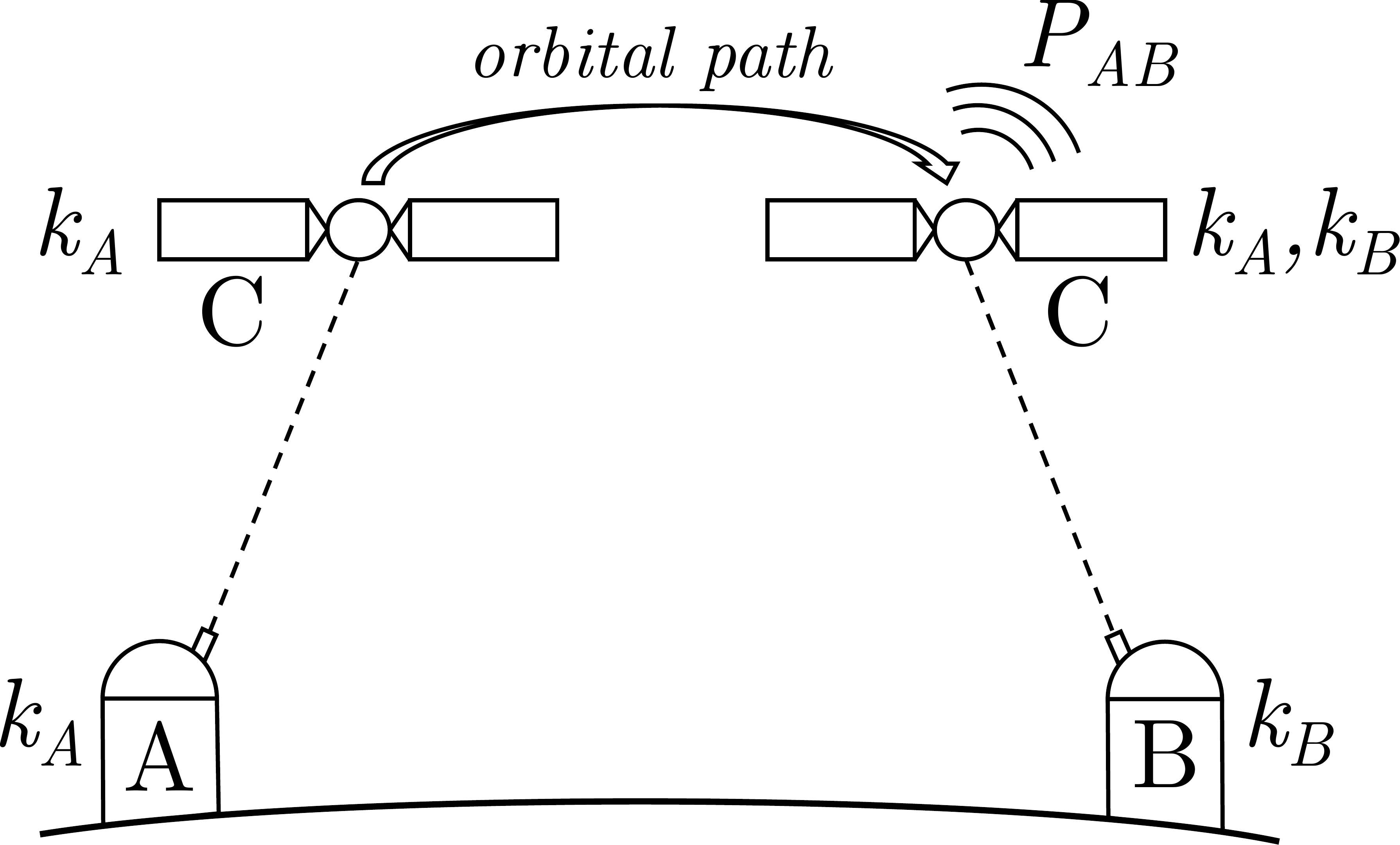}} \\
    \caption{Two possible configurations for satellite QKD. (a) Entanglement-based QKD, in which users A and B can generate a key $k_{AB}$ without the satellite C having any knowledge of the key; and (b) Trusted node QKD, in which users A and B independently generate keys $k_A$ and $k_B$ respectively with the satellite C, at different points along the satellite's orbital path. The satellite can use knowledge of $k_A$ and $k_B$ to broadcast the parity $P_{AB}$, after which A and B can deduce $k_B$ and $k_A$.}
    \label{sat_conf}
\end{figure*}

One possible way of connecting two users using a satellite is the double downlink scenario, depicted in Fig. \ref{sat_conf1}. This particular topology can implement entanglement-based protocols. The satellite (C) is equipped with an entangled photon pair source, and for each pair, it distributes one photon to Alice's OGS (A) and the other one to Bob's OGS (B). A clear advantage of such a configuration is that for many entangled QKD protocols, the entangled photon source need not be trusted \cite{ekert1991quantum}. Unfortunately, both OGS facilities must be simultaneously within sight of the satellite, in practice limiting them to about 1000 km horizontal separation \cite{huttner2022long}. Taken alongside the greatly increased link loss (both photons from a pair must be successfully transmitted), this strategy has to date remained largely of academic interest \cite{yin2020entanglement}.

Requiring only a single optical link, the so-called trusted node approach appears much more feasible to deploy using current technology. The scheme is depicted in Fig. \ref{sat_conf2}. As the first step, the satellite (C) flies over the first OGS (termed Alice) and after establishing an optical link, executes a QKD protocol. Assuming the protocol completes successfully, symmetric keys are generated: one copy stored in the OGS, and the other on board the satellite. At some later time, the satellite flies over the second location (termed Bob) and repeats the same procedure, generating a second pair of keys. At this point, Alice has the key $k_A$, Bob has the key $k_B$, and the satellite has both keys. To connect Alice and Bob, the satellite may disclose (e.g. by broadcast) the parity of the two keys $P_{AB}$, so that Alice and Bob can perform a bitwise XOR operation $(\oplus)$ and obtain the other party's key, which they can now use for cryptographic purposes. The protocol can be formulated as follows:  

\begin{equation}
\begin{split}
    \mathrm{Alice:} \; k_A \oplus P_{AB} = k_A \oplus (k_A \oplus k_B) = k_B. \\
    \mathrm{Bob:} \; k_B \oplus P_{AB} = k_A \oplus (k_A \oplus k_B) = k_A.
    \label{TN}
\end{split}
\end{equation}

For the purposes of this work, it is important to emphasize that once the protocol is complete, inside the satellite the key material is stored classically, thus it is not constrained anymore by the laws of quantum mechanics. The scheme therefore requires Alice and Bob to assume both the security and the honest cooperation of the satellite. External intrusion or malicious action would compromise the security of the whole approach. Thus, insufficient trust in the satellite operator may pose an obstacle in commercializing such services.

Counter-intuitively, the addition of multiple additional trusted nodes can provide a solution to this challenge, provided such nodes are non-cooperative and operated in parallel. While in the introduction we argue that the addition of multiple operators to a trusted node scenario complicates the security of a network, this only holds if the nodes are deployed in a single chain (i.e. in series). In a satellite architecture, a parallel configuration is more natural and provides a mechanism to diversify trust, in which each additional satellite operator provides an effective ``hedge'' against the others.

\begin{table}[]
\caption{Expected launch of satellites in the next two years.}
\begin{center}
\label{table_sat}
\begin{tabular}{|c|c|c|c|}

\hline
\textbf{Name} & \textbf{Nation} & \textbf{Launch} & \textbf{Ref}   \\
\hline
QEYSSat & Canada & 2025-2026 & \cite{qeyssat} \\
\hline
SpeQtre & UK-Singapore & 2025 & \cite{speqtre} \\
\hline 
SpeQtral-1 & UK-Singapore & 2026 & \cite{speqtre} \\
\hline
QUBE & Germany & 2024 & \cite{qube} \\
\hline
QUBE II & Germany & 2024 & \cite{qube2} \\
\hline
ROKS & UK & 2024 & \cite{roks} \\
\hline
NanoBob & France & 2024 & \cite{NanoBob} \\
\hline
Hub IOD & UK-Netherlands & 2024  & \cite{iod} \\
\hline
Eagle-1 & EU & 2024 & \cite{Eagle} \\
\hline

\end{tabular}
\end{center}
\end{table}

\section{Diversification of trust}

\begin{figure*}
    \centering
    \includegraphics[width=0.7\textwidth]{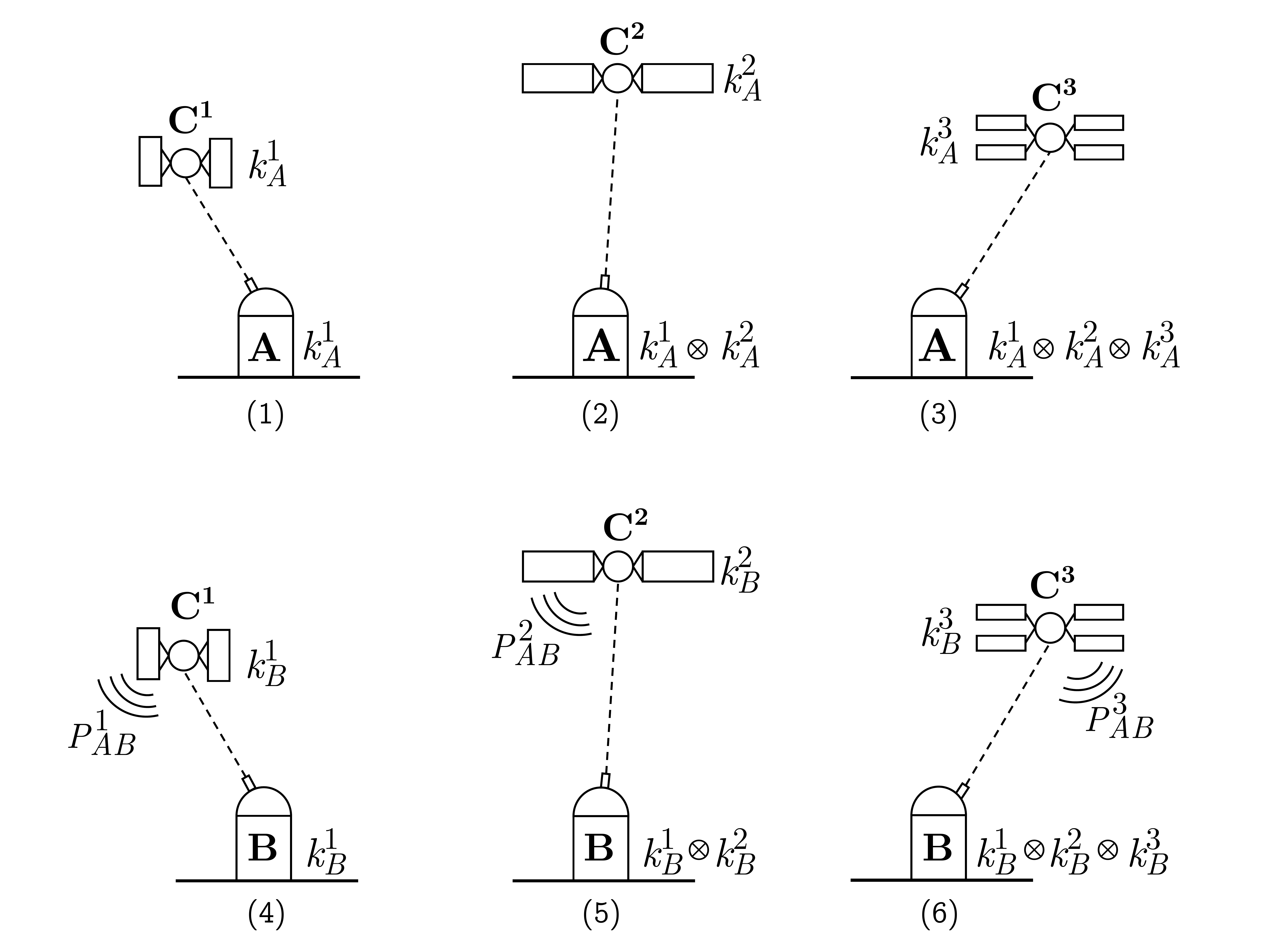}
    \caption{Representation of the parallel trusted node scheme for n=3. The first satellite $C^1$ generates the pairs of sub-keys $k_A^1$ and $k_B^1$, and it announces their parity (Figs. 1 and 4). Similarly for the second satellite $C^2$ (Figs. 2 and 5) and the third satellite $C^3$ (Figs. 3 and 6). The two OGSs first obtain the final keys through the bitwise operation of their sub-keys and then retrieve the other's party key using the publicly announced parities. The order in which the communications occur is irrelevant.}
    \label{fig:PTN}
\end{figure*}

The proposed parallel TN QKD (Fig. \ref{fig:PTN}) extends the elementary trusted node approach described in Eq. \ref{TN}. Utilizing the diverse QKD satellites deployed by various operators, it is conceivable for OGSs A and B to establish independent QKD sessions with distinct satellites. The approach is parallel because the result of each sub-protocol is not used during the other sub-protocols. There is no constraint required on the order in which OGSs communicate with the participating satellites. After independently establishing key pairs with $n$ different satellites, Alice and Bob obtain $n$ distinct sub-keys. Subsequently, the bitwise XOR operations of these sub-keys and the publicly announced parities from all the satellites generate the final cryptographic keys. The approach can be stated as:

\begin{equation}
\begin{split} 
    \mathrm{Alice:} \; \bigoplus_{i=1}^n k_A^i \oplus \bigoplus_{i=1}^n P_{AB}^i  = \bigoplus_{i=1}^n k_B^i. \\
    \mathrm{Bob:} \; \bigoplus_{i=1}^n k_B^i \oplus \bigoplus_{i=1}^n P_{AB}^i = \bigoplus_{i=1}^n k_A^i,
\label{Eq_parallel_TN}
\end{split}
\end{equation}

where $k_A^i$, $k_B^i$, and $P_{AB}^i$ are the sub-keys and the parity generated using the i-th satellite, and $(\bigoplus_{i=1}^n)$ is the bitwise XOR operation of n bitstrings. So, we propose to use a key derivation algorithm that produces the final cryptographic key from all the parallel sub-keys shared via independent TNs. A simple and information-theoretic secure method for doing this is described in Eq. \ref{Eq_parallel_TN}. The lack of knowledge about any individual elementary key leaves the adversary without any information about the final key. This results in substantially weaker assumptions about our trust in the QKD satellite vendors. The eavesdropper must access all parallel independent TN key exchanges to have information on the users' final keys. This is in principle possible, but from a practical point of view, Eve has to have information access to several different satellites that are owned, assembled, and managed by various companies or governments, which is a significantly greater challenge.

Moreover, the parallel TN concept helps mitigate hardware vulnerability issues. Unlike the idealized QKD models, the actual hardware can exhibit non-ideal behavior, and this can lead to vulnerabilities given malicious signals injected into the quantum channel~\cite{gisin2006trojan,lydersen2010hacking}.
In the past, even commercial implementations have been shown to be vulnerable to such hardware attacks ~\cite{gerhardt2011hacking,xu2010experimental,zhao2008quantum}. Although modern QKD implementations are much more durable, there are non-zero chances that new attack surfaces may be discovered impacting specific hardware realizations. The proposed TN approach with its trust diversification diminishes the probability of successful hardware exploitation, helping to mitigate the risk of a global security breach. It is apparent that even when employing the same QKD protocol, different vendors are prone to implement slightly diverse hardware configurations and software algorithms. Under parallel TN conditions, the full-scale breach of one QKD session does not jeopardize the integrity of the overall global key security, highlighting the robustness of this model.

The possibility of secure communication using multiple TNs strongly depends on the number of satellites available to the users, and its feasibility is motivated by the rapid expansion of the satellite QKD market, evidenced by the growing number of operators (governments and private companies) entering into it. Acknowledging the strategic importance of secure quantum communication infrastructure, entities worldwide firmly invest in satellite QKD missions to bring quantum security to real-life applications such as finance, defense, industry, critical infrastructure, and administration \cite{de2023satellite}. Tab. \ref{table_sat} provides a non-exhaustive list of upcoming satellite launches for quantum communication expected in the next two years.

\section{Versatile optical ground stations}

The realization of such a user-favorable scenario relies on the \emph{versatility} of the available OGS facilities, which must be capable of communicating with several different satellites, ideally built and operated by diverse teams. Consequently, the adaptability and multifunctionality of an OGS emerge as crucial factors in enabling the parallel TN approach. The downlink QKD scenario is assumed below. 

A standard component of any OGS is a telescope with a fine acquisition and tracking system, which aims to collect light and send it to a measurement apparatus. While different satellites can use different protocols and encoding methods, the collection of light and the correction of angular deviations are universal tasks which in practice are achieved similarly. 

Enabling satellite QKD requires two distinct downlink optical channels: the quantum channel (used to deliver photonic quantum states), and the beacon channel which serves the purpose of acquisition and tracking. Optionally, other channels may be added to the system, e.g. for classical data transfer. Together with the beacon channel, they will be referred to as ``classical channels'' below.

To date, most satellite QKD implementations or proposals use quantum channels in the visible spectrum. This is due to the combination of the atmospheric transmission spectrum and the availability of compact, high performance single-photon detectors at these wavelengths. While advances in superconducting nanowire detector technology are expected to extend the available spectrum, today the most commonly adopted detectors are silicon-based avalanche photodiodes (SPADs). These feature optimal performance in the spectral range of 500 -- 850 nm. Atmospheric losses further constrain today's systems to 750 -- 850 nm, dominated by Rayleigh scattering that decreases at longer wavelengths \cite{andrews2005laser}. This relatively narrow range can simplify OGS receiver design, as the whole band may be efficiently split from the incoming beam and routed into the quantum optics package. In the future, we anticipate the community may look to adopt longer wavelengths, for example, the telecom bands between 1300 and 1600 nm. A notable reduction in Rayleigh scattering loss is observed at these wavelengths, albeit at the tradeoff of increased diffraction-limited beam divergence. Moreover, this option is often proposed in the context of daylight QKD missions~\cite{avesani2021daylight}, where the reduced solar radiation at these wavelengths can result in improved signal to noise.

Conversely, when considering classical channels, the wavelength choice is less constrained and, thus, more diverse, spanning roughly from 500 to 1610~nm~ \cite{kaushal2015free}. At present, there is almost no consensus on which wavelengths to adopt for satellite QKD beacons. As such, it is this point at which concerns of versatility enter into the OGS design. A realistic near-universal detection strategy can be constructed by designing around two pointing error sensors with silicon (for $\lambda \lesssim 1000$~nm) and InGaAs (for $\lambda \gtrsim 1000$~nm) sensors. This could cover most wavelength choices for the satellite beacon, and would provide the required flexibility to the OGS facility.

Outside of the QKD community, significant attention is directed toward standardizing free-space optical communications for satellite applications. In particular, there are CCSDS standards for 1064~nm and C-band communications~\cite{ccsds-standards} as well as the SDA optical communication terminal standard~\cite{sdaoct-standard}, which also specifies C-band communication.  Although the available standards do not consider the presence of QKD channels (and so do not cover the whole system design), they may be valuable as a reference. Future standardization efforts in the QKD satellite field are essential for enabling a versatile approach to OGS designs.

A versatile receiver should be designed to divide the input light of different wavelengths into different sectors, each equipped with optical elements and detectors specific to that particular range. Moreover, OGSs should support uplink beacon lasers to facilitate tracking satellites, potentially combined with classical (optical) data channels. This is not usually an engineering challenge, as these lasers typically do not pass through the main telescope aperture. The classical beams are emitted from much smaller transmission optics, hence several transmission modules may be mounted on the telescope without significantly increasing cost or complexity. Alternatively, the transmission module can be designed for a broad wavelength range, facilitating rapid reconfiguration between several satellite missions. The transmission beams can also be delivered over optical fiber. If this is the case, reconfiguring the OGS may be implemented via swapping the feeding fibers without duplicating transmission optics.

Several launches of satellite missions dedicated to quantum key distribution are planned for the near future. It is expected that a similar proliferation of OGSs with some degree of versatility will be implemented to integrate the vision of a future global quantum network. In planning an OGS, one requires both a receiver module that can detect a broad spectrum of radiation and a suitable set of transmitter modules that can emit in a wide range of wavelengths, and we hope that our notes here can serve as some reference for future mission planners.

\section{Conclusions}
Satellites are a crucial element for achieving quantum communications at a global scale. Without the availability of quantum repeaters and quantum memories, the trusted node approach, where the users must trust both the satellite hardware and its operator, is required to cover intercontinental distances. 
We have proposed and elaborated a novel solution to realize satellite QKD with less stringent assumptions than the elementary TN scenario, making use only of currently available technology. Our approach comes at the cost of interacting with more than one satellite, which carries a cost in terms of the resources needed at the receiver. However, when considering the number of new missions that are projected to launch in the near future, we believe this proposal presents an attractive and indeed sensible way to advance the adoption of satellite quantum communication technology without putting full emphasis on a single operator. In general, independent QKD satellite missions with diverse hardware implementations are a more attractive, safe, and robust configuration than a single global satellite operator relying upon a single type of QKD-enabled fleet.

Lastly, we outlined the core OGS features to realize the proposed approach. Versatility, as the paramount ability to connect the station with several different satellites, is the required property to overcome the basic trust issue of the elementary TN QKD, and to provide full integration into the global quantum network.

\section*{Acknowledgements}

We thank Anton Trushechkin, Rodrigo S. Piera, and Yury Kurochkin for the helpful discussions.

\bibliography{biblio}

\end{document}